# Will Remote Work Drive a New Wave of Suburbanisation in Poland? Analysing the Relocation Preferences of Polish Office Employees

Sławomir Kuźmar, Beata Woźniak-Jęchorek, David Bole

**Abstract:** This study assesses how the growing availability of working from home (WFH) shapes office employees' preferences to move to the suburbs and pinpoints the socio-economic factors that drive those intentions. We focus on Poland, where the housing market is shaped by exceptionally high home-ownership rates and specific suburbanisation patterns. We surveyed city-dwelling office employees (living in municipalities of 100,000 or more) to gauge their willingness to relocate. Logistic-regression estimates then linked those intentions to respondents' demographics, job attributes, commuting patterns, and self-reported productivity shifts under WFH. The study tests three mechanisms. Commuting cost is proxied by travel mode and one-way time; life-course triggers by age, children, and tenure; and job-demands/resources by self-rated productivity under WFH. Sector and city size serve as contextual controls. Linking variables to theory in this way clarifies how the forthcoming results adjudicate among competing explanations. The results indicate that age, commuting mode, self-assessed productivity changes, and employment sector (private versus public) markedly influence the likelihood of considering a move to the suburbs in response to remote-work options. Contrary to expectations, household size, measured by number of children, does not play a significant role. Overall, the evidence suggests that remote work, especially in hybrid form, could become an additional catalyst for suburban expansion in markets characterised by scarce affordable rentals and a strong preference for home ownership, such as Poland.

## Introduction

Across numerous countries, the post-COVID spread of work-from-home (WFH) [1] and other flexible arrangements has reinforced the trend of migration from downtown districts to outlying suburbs, as reduced commuting needs help alleviate the spatial constraints that have historically limited residential mobility and make suburban living an even more attractive option. Pioneering studies on telecommuting anticipated this outcome (Lund & Mokhtarian 1994; Ory & Mokhtarian 2006; Ettema 2010; Kim 2016), and an expanding body of post-pandemic research now documents the pattern in North America, Western Europe, and East Asia alike (e.g., Liu & Su 2021; Ramani & Bloom 2021; Gupta et al. 2022; Stefaniec et al. 2022; Huang et al. 2023; Gokan et al. 2023; Kim & Long 2024; Jansen et al. 2024; Sa & Shen 2025, Hostettler Macias et al. 2025; Wächter & Kramer 2025; Marinos et al. 2025).

[1] The terms "remote work," "teleworking," "telework" and "working from home" all relate to work's spatial distribution and often overlap. Teleworking can include various offsite locations and may not always replace commuting. This research focuses on remote work as an arrangement where essential job duties are performed away from the primary worksite, serving as a substitute for working at home.

Most of that literature, however, comes from rental-heavy housing systems where relocation costs are low and short leases are common. In post-socialist Poland, by contrast, mass privatisation in the 1990s pushed owner-occupation above 75.4 per cent (Eurostat, 2024). In markets such as Poland, where institutional rentals are scarce, this centrifugal pull is thus further amplified because households cannot simply switch leases; moving generally means selling one owned dwelling and buying another. Equity locked in owner-occupied dwellings, lengthy selling times, and high transaction fees suppress residential mobility, especially for older cohorts who have already repaid their mortgages. Younger owners may still absorb those costs if WFH lets them swap a small inner-city flat for suburban space, but older owners face steeper financial and psychological barriers. WFH therefore sets up a natural experiment: does the gain in daily flexibility outweigh these ownership frictions, or do transaction costs still anchor households to the core?

To test these ideas, we focus on Poland's larger cities (≥ 100,000 residents) and ask:

- RQ1: Does access to WFH alter residents' inclination to relocate to the suburbs?
- RQ2: How are such intentions shaped by (a) socio-economic and job attributes, (b) commuting experience, and (c) self-reported productivity changes when working remotely?

We fielded an online survey that gathered 639 responses from city-based office employees, including on-site staff whose tasks could feasibly shift online (Dingel & Neiman 2020). The analysis targets three explanatory channels through a binary logit model: commuting costs, captured through usual travel mode and single-trip duration; life-course triggers, represented by age, parental status, and housing tenure; and job demands versus resources, gauged via respondents' self-assessed productivity when working remotely. Employment sector and core-city population enter as background controls. Aligning each variable set with a specific theoretical mechanism makes it clear how the empirical results will discriminate among the rival explanations.

The results reveal a selective, rather than universal, decentralisation impulse. Younger workers, car passengers, private-sector employees, and those reporting productivity gains are markedly more likely to consider a move, whereas family size and city population have no statistically discernible effect. These patterns imply that hybrid WFH could become an additional driver of suburban expansion in owner-heavy markets, such as Poland's, where rental alternatives are scarce.

The remainder of the paper proceeds as follows: Section 1 reviews theoretical and empirical work on residential relocation in the context of remote work; Section 2 outlines Poland's housing and suburbanisation context; Section 3 details the data and modelling approach; Section 4 presents and interprets the findings; and Section 5 concludes with policy implications and avenues for future research.

## 1. Remote Work and Residential Location Choice: Theoretical Factors

Residential choice is a long-standing topic in economics, geography, and the social sciences because it shapes both household welfare and urban form. The modern analytical lineage begins with von Thünen's *Isolated State* (1826), which showed that the trade-off between transport costs and land rent determines where people live and work. Neoclassical equilibrium thinking later framed location as a utility-maximising decision under budget constraints, an idea formalised by Tiebout (1956), who argued that households "*vote with their feet*" for preferred mixes of taxes and public goods.

Subsequent work added a household lens. Mincer's family-migration model (1978) emphasised how dual careers and kin ties raise the cost of moving, while Mulder and Hooimeijer (1999) embedded relocation in the life course: events such as partnership, childbirth, or home purchase both trigger and constrain moves.

Urban economists complemented these perspectives with spatial-equilibrium models. Alonso's monocentric city (1964) predicted a gradient of falling land prices and densities away from the central business district (CBD), with households trading cheaper space for longer commutes. Later extensions showed that income growth, transport investment, and – more recently – telework can tilt that balance. Ota and Fujita (1993) demonstrated that parameter shifts can yield either compact or dispersed cities, and empirical studies (e.g., Lund & Mokhtarian 1994; Nagurney et al. 2003; Rhee 2009) confirm that telecommuting shortens trip frequency but often pushes households towards more distant suburbs, reshaping metropolitan structure.

Although early models implied that telecommuting would reshape residential choice, empirical evidence remains mixed. Ory and Mokhtarian (2006) observed that some telecommuters moved nearer to their jobs, yet those who relocated farther afield often adopted teleworking only afterward. Ettema's (2010) latent-class analysis likewise detected no systematic relocation effect, though it did reveal groups that differed in their sensitivity to commuting distance. Employing path analysis, Kim (2016) concluded that job location, not the option to telecommute, was the decisive factor in both workplace and homesite selection.

The pandemic, however, shifted the debate. Remote work surged during COVID-19, and although full-time telework has since ebbed, hybrid schedules remain far more common than before 2020 (Aksoy et al. 2025; The Economist, 2025). Post-pandemic studies generally point to renewed suburbanisation as lower commuting frequency makes peripheral living more viable (Brueckner et al. 2023; Akan et al. 2024). Some authors report stronger preferences for suburban or even rural settings (Gong et al. 2024), whereas others find little evidence of a countryside exodus (Neumann et al. 2022). Van Acker et al. (2024) note that greater residential satisfaction under WFH may dampen large-scale moves, yet when office attendance is still required, commute distances can lengthen (Hostettler Macias et al. 2022).

Patterns also vary by local context. In several U.S. metros, a "doughnut" effect emerged, with downtowns retaining strength relative to inner suburbs, though weaker regional economies show different trajectories (Ramani & Bloom 2021; Chun et al. 2022; Lee & Huang 2022). Gong et al. (2024) provide strong evidence that remote work, income, and housing supply jointly push equilibrium toward suburbia or fewer trips altogether. By contrast, Sa and Shen (2025), using Seattle's 2022 Commute Survey, find that full- or hybrid-remote employees are likelier either to stay put or to relocate farther away, not necessarily to move closer. Because similar factors – work pattern, income, age, household size – drive both "stay" and "move farther" decisions, the authors conclude that remote work expands spatial flexibility without uniformly propelling households outward.

## 2. Distinctive Features of the Housing Market and Suburbanisation in Poland

Poland's housing market combines two features that set it apart within Europe: an exceptionally high rate of owner-occupation and a rapid, largely uncoordinated wave of suburbanisation that followed the political-economic transition of 1989. According to Eurostat, roughly 75.4 percent of Polish households occupy their dwellings on an ownership basis, compared with an EU average of about 44.2 percent in 2024 (see Fig. 1). Additionally, figure 2 shows that Polish households have, on average, only 1.2 rooms per person, well below the EU mean of 1.8. This relative crowding in urban housing likely fuels the migration toward the suburbs, where larger and more spacious homes are easier to find.

Unusually high homeownership rates in post-socialist countries can be explained by several factors. One of them is the massive privatisation of dwellings from the housing stock constructed or taken over by the state in socialist times (Buckley & Tsenkova, 2001). In all CEE countries, housing units were generally sold at prices well below their market value, incorporating explicit discounts (Broulíková & Montag, 2020). In Poland, the discount was typically over 80% (Skiba, 2005) and could amount to up to 95% of the market value (Lis & Zwierzchlewski, 2015). Country-specific factors also played a certain role. For example, Poland had a relatively high number of owner-occupied single-family houses even under socialism, because a significant share of land (mostly in rural areas) was private (Radzimski, 2014). The constitutional obligation placed on public authorities to facilitate home purchase further entrenched ownership as the dominant tenure norm reinforced the cultural expectation that secure housing is synonymous with owning rather than renting.

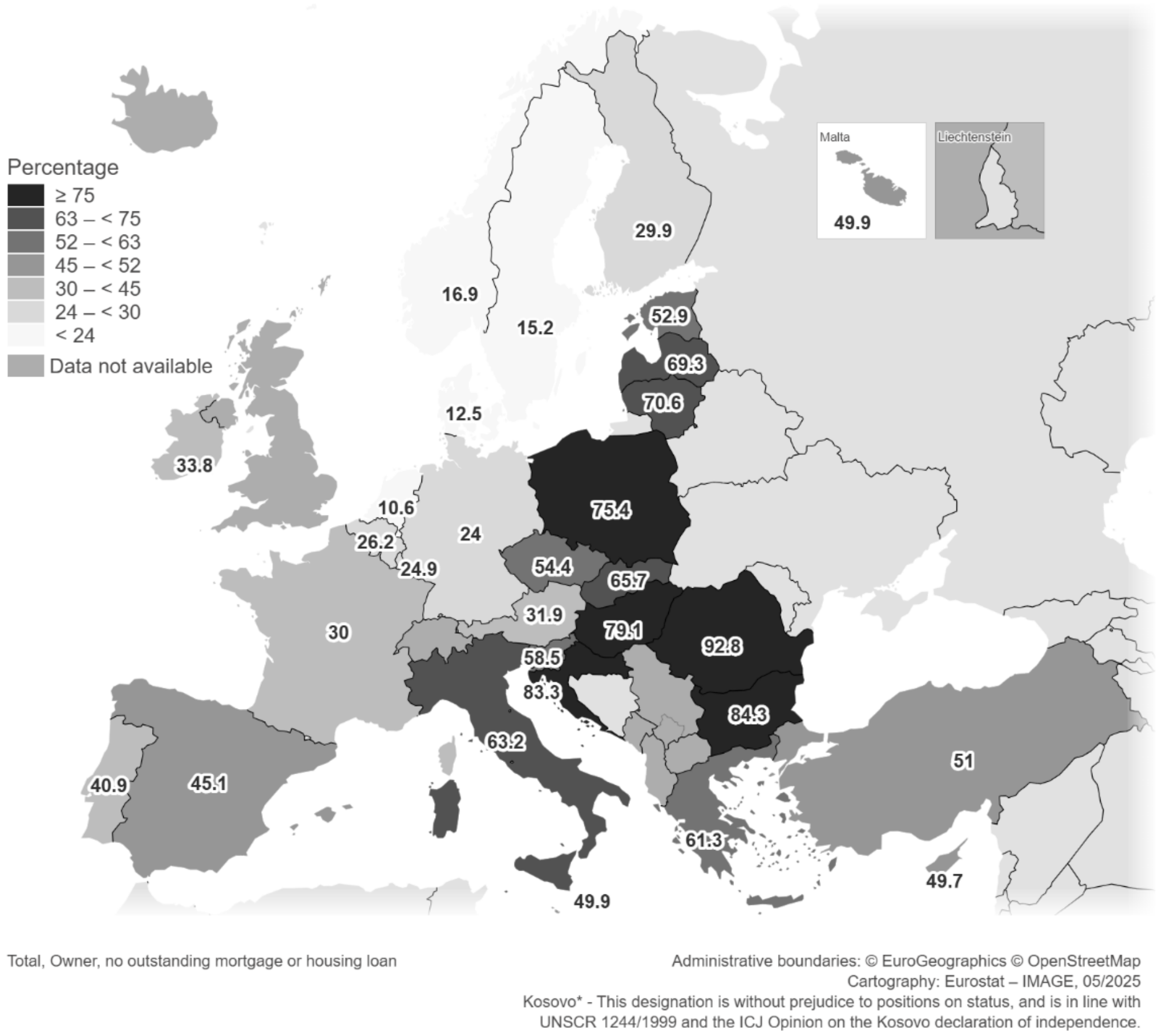

**Figure 1. Share of people living in households owning the place (without mortgage) in 2024**

Source: *own elaboration based on Eurostat data.*

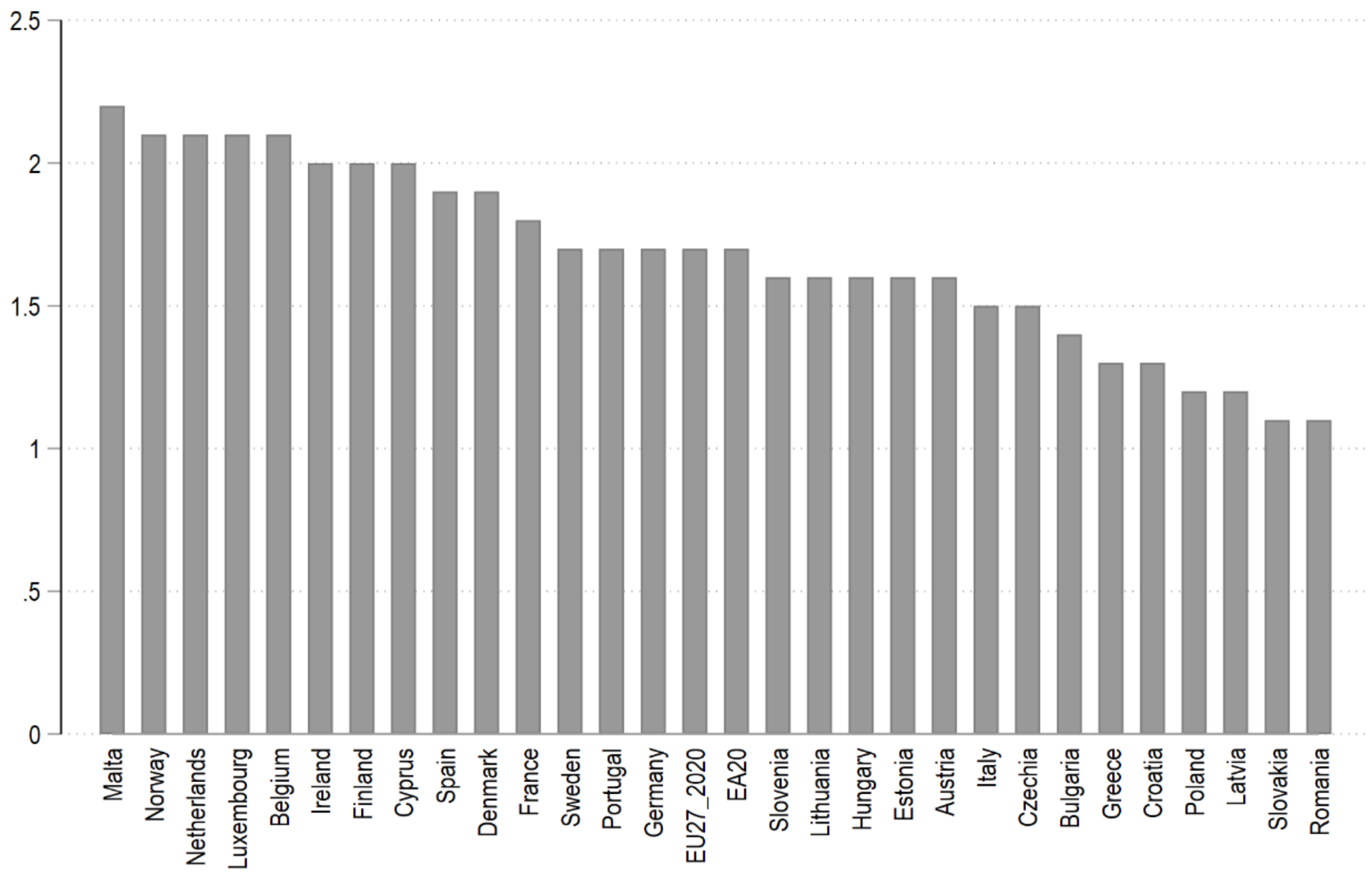

**Figure 2. Average number of rooms per person in 2024**

Source: *own elaboration based on Eurostat data.*

The corollary of this transformation is an unusually small institutional rental sector, which has important macroeconomic consequences. High ownership rates tend to lower household mobility because transaction costs and psychological attachment to property discourage frequent moves; Glocker and Plouin (2016) have linked this structural immobility to Poland's relatively sluggish internal migration despite marked regional differences in labour demand. Limited rental supply also delays residential independence: 51% of Poles aged 25–34 still resided with their parents in 2022 (Polish Economic Institute, 2024).

The interplay of widespread homeownership and relatively cramped urban dwellings has also shaped Poland's distinctive pattern of suburban growth. While Poland exhibits the post-socialist pattern of population drift towards the fringe (Koj 2020), the process follows its own logic. Kajdanek (2011) argues that the scale and intensity of Polish suburbanisation remain below those observed in Western Europe, yet its rate of expansion has been remarkably rapid since the early 1990s, visibly reshaping suburban landscapes and lifestyles. Most new housing estates lie only a few kilometres from the urban core (Palak 2016), often abutting prefabricated blocks; the result is a heterogeneous mosaic of single-family villas, high-rise flats, village cottages, small workshops and remnant farmland – an urban–rural blend rarely seen in Western suburbs (Więcław-Michniewska 2006). Socially, the Polish suburban population differs from its Western counterpart, reflecting post-transition stratification (Kajdanek 2012; Ouředníček 2007). Suburban homes are typically larger than compact urban flats, which often lack sufficient space – a contrast that further fuels the appeal of suburban living (Więcław-Michniewska 2006). Low residential density and dispersed building patterns even draw comparisons with North American suburbia (Palak 2016).

The geography of sprawl is also uneven. Dawid (2021) shows that growth is strongest around the largest metropolitan areas; Mayer and Szmytkie (2014) documented that these metropolitan belts were the pioneers of rapid out-migration, later emulated by medium and small towns (Zborowski & Raźniak 2013). The OECD (2022) calculates that between 1990 and 2015, core cities lost 8.6% of their population, while semi-dense areas (suburbs) grew by 5.2%. Moreover, 43% of residents of Functional Urban Areas (FUA) live in the commuting zones, far above the OECD average (25%), which is an additional indicator of suburbanisation.

Several interacting forces underpin this dispersal. First, shortages of modern, spacious dwellings in city centres made peripheral plots attractive to an expanding middle class eager for single-family housing. Second, abundant agricultural land at urban fringes was cheaply converted to residential use under a permissive and fragmented planning regime. Third, a high growth rate of passenger cars per 1000 inhabitants – from 495 vehicles in 2012 to over 584 in 2022 (Eurostat 2024) reduced the effective cost of longer commutes and enabled daily travel from dispersed settlements. The outcome is a settlement pattern characterised by low-density sprawl, infrastructural deficits, and longer journey times.

## 3. Work-from-Home and Residential Preferences in Poland: Empirical Verification

Synthesising the foregoing literature, our empirical specification targets three explanatory channels. (1) Commuting costs, central to monocentric-city models, are proxied by respondents' usual travel mode and single-trip commute time, together standing in for the monetary and time 'price' of distance. (2) Life-course triggers, highlighted in mobility and family-migration studies, are operationalised via age categories, household structure (presence of children), and tenure, variables that modulate relocation propensity at different family stages. (3) Job-demands/resources, drawn from occupational-psychology frameworks, are gauged by each respondent's self-assessed change in productivity while working from home, a synthetic measure of how extra demands (e.g., work-family conflict) are offset by resources (e.g., separate workspace, social support). The sector of employment and the population size of the core city enter as contextual controls that may modulate the influence of these three mechanisms.

### *Dataset Attributes and Response Statistics*

To understand how WFH influences the desire to move from a big city to the suburbs in Poland – and how that desire depends on workers' socio-economic profiles, job circumstances, commuting habits, and perceived productivity – we designed an online questionnaire for office employees in Polish largest urban centres (population > 100 000).

We focused on larger cities, where remote and hybrid work are more prevalent (Kapitsinis 2025; Qian & Su 2025) and relocation dynamics are more pronounced. Suburbanisation in Polish metropolitan areas correlates strongly with town size and economic development (Gałka & Warych-Juras 2018), with major cities experiencing higher population growth than smaller ones, intensifying residential pressure (Gnat, 2024).

The survey, administered through the Ariadna national web panel, was conducted from 4 to 7 June 2024, by which time attitudes toward remote work had stabilised after the pandemic. The sample consisted of 639 respondents and was designed to achieve a confidence level of 95%, a maximum margin of error of 0.04%, and a fraction size of 0.5, considering the reported 5.1 million employed office workers in the national economy in Poland. We are not able to verify the representativeness of our sample specifically for office workers residing in larger cities, as no official statistics exist for this particular subgroup. However, it seems that our sample aligns reasonably well with general demographic distributions. For example, according to Eurostat (Eurostat, 2025), the share of women and men among urban residents is approximately 51.7% and 48.3%, respectively – figures that correspond closely to the gender composition observed in our dataset[2].

The questionnaire featured single- and multiple-choice items. Socio-demographic measures included gender, age, city of residence, education, and number of children under 18. Employment variables covered economic sector, ownership type (public/private), role in organization, and years in the current post. A full list of variables is provided in Table 1.

[2] Although a direct benchmark for Poland's city-based office workforce is unavailable, the near-parity gender split suggests the panel is broadly plausible; nonetheless, the provider's recruitment algorithm draws disproportionately from the largest metropolitan areas – those with population over 500k inhabitants – so the sample may over-represent employees with greater exposure to telework and higher housing costs. This potential "metro-area bias" could inflate the overall propensity to relocate and should be borne in mind when generalising the results to smaller urban systems.

**Table 1. Respondent's socio-demographic and employment characteristics**

| Socio-economic characteristics | | Freq. | Percent | Cum. |
|---|---|---|---|---|
| **Gender** | Women | 327 | 51.2 | 51.2 |
| | Men | 310 | 48.5 | 99.7 |
| | Non-binary | 1 | .2 | 99.8 |
| | I prefer not to disclose | 1 | .2 | 100.0 |
| **Age** | 25-34 years | 205 | 32.1 | 32.1 |
| | 35-49 years | 267 | 41.8 | 73.9 |
| | 50-54 years | 58 | 9.1 | 82.9 |
| | 55-60 years | 53 | 8.3 | 91.2 |
| | 61-65 years | 31 | 4.9 | 96.1 |
| | +66 years | 25 | 3.9 | 100.0 |
| **Place of residence** | City from 100.000 to 199.000 inhabitants | 135 | 21.1 | 21.1 |
| | City from 200.000 to 500.000 inhabitants | 161 | 25.2 | 46.3 |
| | City up to 500.000 inhabitants | 343 | 53.7 | 100.0 |
| **Education** | Primary/Middle School | 1 | .2 | 0.2 |
| | Vocational | 5 | .8 | 0.9 |
| | Secondary | 73 | 11.4 | 12.4 |
| | Post-secondary/Vocational College | 45 | 7.0 | 19.4 |
| | Higher Education - Bachelor's | 92 | 14.4 | 33.8 |
| | Higher Education - Master's Equivalent | 423 | 66.2 | 100.0 |
| **Number of children under 18 in the household** | No kids | 379 | 59.3 | 59.3 |
| | One | 151 | 23.6 | 82.9 |
| | Two | 96 | 15.0 | 98.0 |
| | Three | 9 | 1.4 | 99.4 |
| | Four | 3 | .5 | 99.8 |
| | Up to five | 1 | .2 | 100.0 |
| **Role in organisation** | Office administrative employee | 166 | 26.0 | 26.0 |
| | Junior specialist | 95 | 14.9 | 40.8 |
| | Senior specialist | 237 | 37.1 | 77.9 |
| | Director/Manger | 99 | 15.5 | 93.4 |
| | Company owner | 42 | 6.6 | 100.0 |
| **Numbers of years in the role** | Less than a year | 57 | 8.9 | 8.9 |
| | 1-3 years | 139 | 21.8 | 30.7 |
| | 4-7 years | 171 | 26.8 | 57.4 |
| | 8-10 years | 63 | 9.9 | 67.3 |
| | More than 10 years | 209 | 32.7 | 100.0 |
| **Ownership structure** | Public | 216 | 33.8 | 33.8 |
| | Private | 423 | 66.2 | 100.0 |
| **Sector** | Industry | 43 | 6.7 | 6.7 |
| | Technology and computer science | 84 | 13.1 | 19.9 |
| | Finance and banking | 61 | 9.5 | 29.4 |
| | Education | 64 | 10.0 | 39.4 |
| | Healthcare | 36 | 5.6 | 45.1 |
| | Trade and services | 119 | 18.6 | 63.7 |
| | Transport and logistics | 39 | 6.1 | 69.8 |
| | Media and entertainment | 20 | 3.1 | 72.9 |
| | Other | 173 | 27.1 | 100.0 |

Note: Due to the low number of observations, responses from the "Non-binary" and "I prefer not to disclose" were excluded from further analysis. As a result, the total number of observations used in the analysis was 637.

The survey asked participants how far from the city centre they would live if remote work were feasible, their present commuting mode, and their past WFH experience, including the preferred share of workdays spent at home. We also recorded self-assessed changes in productivity and quality of life under remote work and invited respondents to name their primary motives for contemplating relocation.

In the collected sample, 55.7 % of office employees still work entirely on-site, while 13.3 % are fully remote; the remainder split their time between home and office one to four days per week. When asked about their preferred arrangement, enthusiasm for remote work rises markedly: 21.4 % would like to work exclusively from home, 20.7 % would choose to stay home for more than two days per week, and 19.1 % favour exactly two remote days.

The core survey item examined willingness to relocate. Among respondents who can work remotely, one in two (50.4 %) would consider moving from the city to a suburban location because of WFH opportunities, and nearly half of that group (44.4 %) would limit the move to within 50 km of the city centre (Fig. 3).

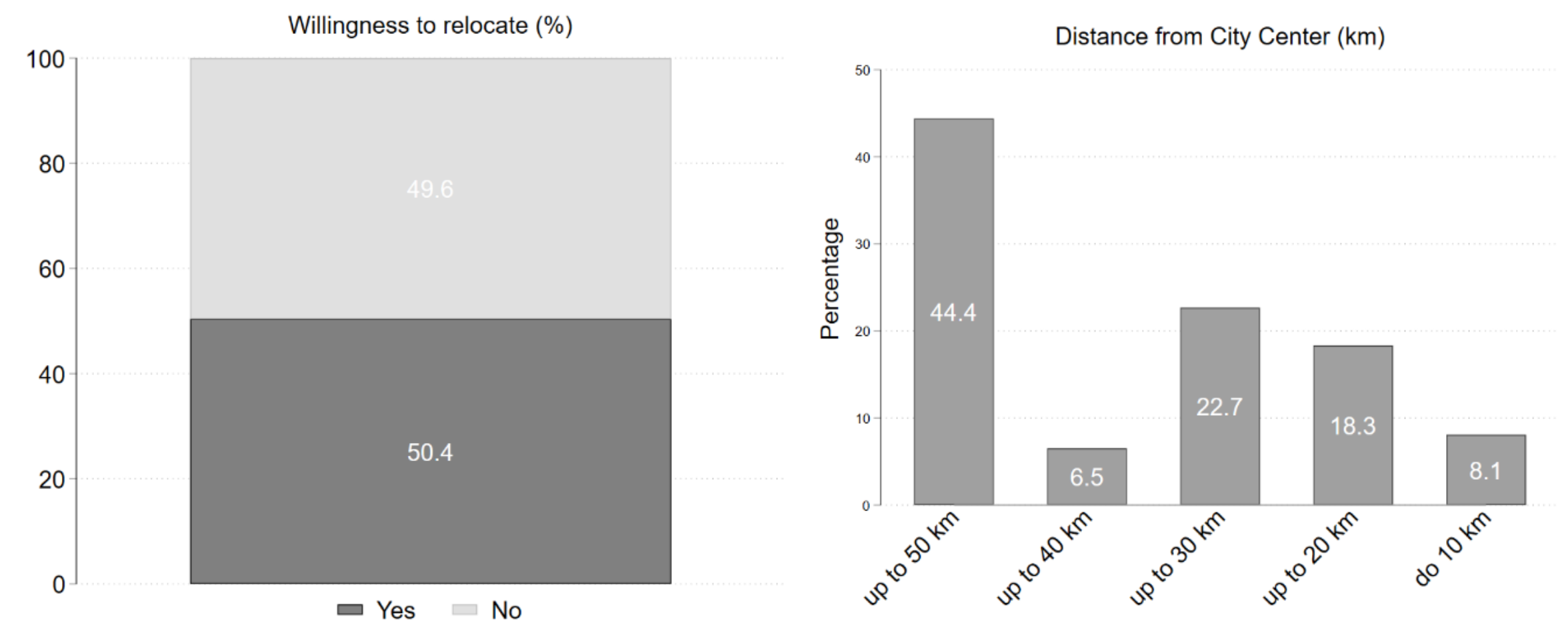

**Figure 3. Distribution of Preferences for Relocation and Preferred Distance**

Note: Responses to the questions: "If you had the option to work remotely, would you consider moving to the suburbs?" and "If you had the option to work remotely, how far from the city center would you be willing to live in the suburbs?"

Source: *own elaboration based on survey data.*

Over 87% of respondents see 45 minutes as the longest acceptable commute. When commuting was required only occasionally, however, tolerance for a longer journey rose sharply: the share willing to travel more than an hour each way jumped from 2.2% to 12.4% (Fig. 4). Most respondents reach the office by driving (56.7%), followed by public transport (41.2%), walking (19.6%), cycling (13%) and riding as a car passenger (9.1%). Rail accounts for 4.9 %, while a further 5.3 % rely on micro-mobility options such as e-scooters or mopeds.

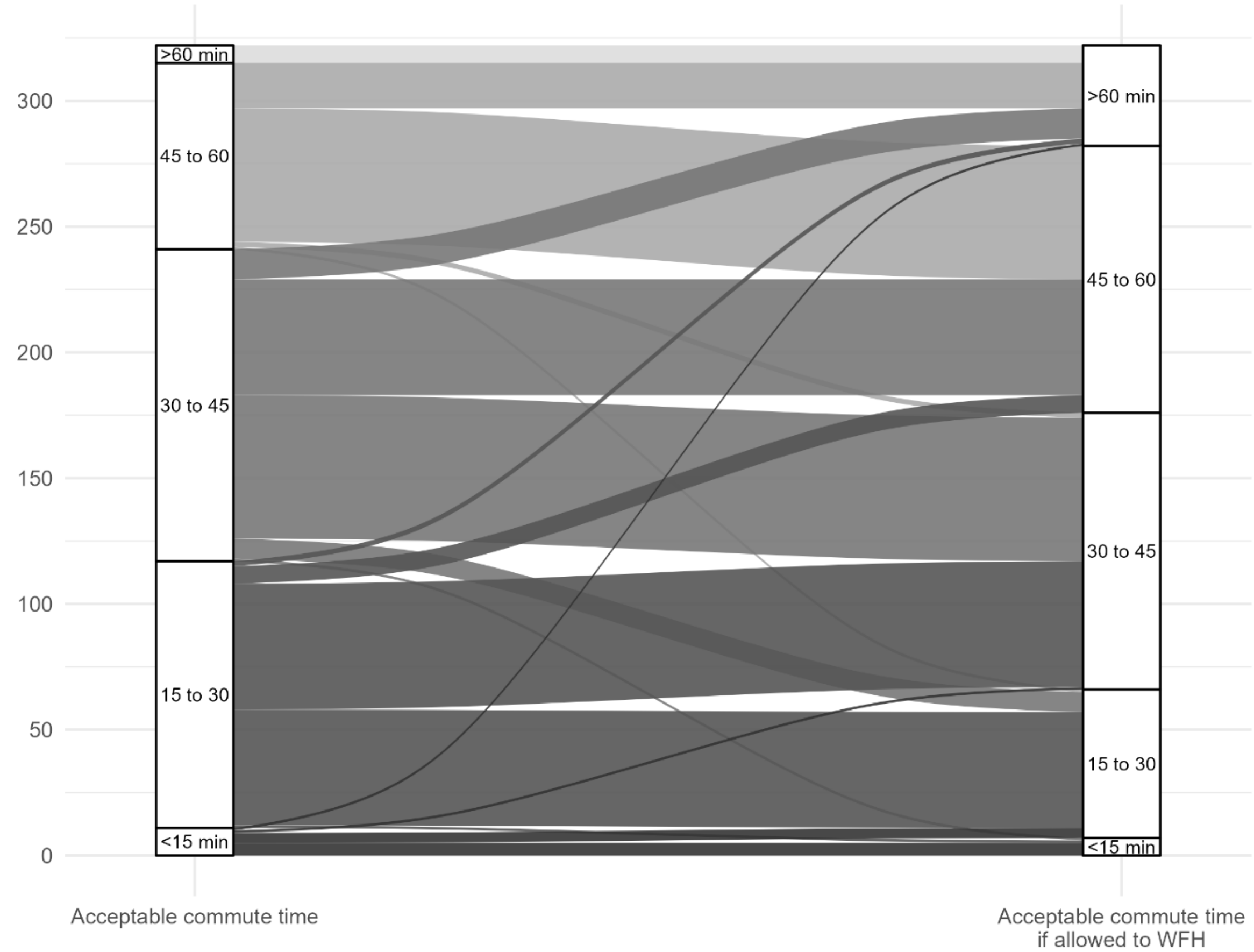

**Figure 4. Employees' acceptable commute time without and having WHF possibilities**
Source: *own elaboration based on survey data.*

When respondents were asked why they might move to a suburb, three benefits dominated their answers: greater proximity to nature (74.5 %), the prospect of more living space at a lower price (72.4 %), and overall lower living costs outside a large city (60.2 %) (Fig. 5). Factors such as lighter traffic, nearness to family and friends, or better family-oriented infrastructure ranked considerably lower.

Relocation intentions also appear to hinge on how people feel remote work affects their daily lives. As Fig. 6 shows, most participants viewed WFH positively: 45.2 % felt more productive, and 39.0 % reported a better work–life balance.  Only a small minority saw major declines in either dimension.

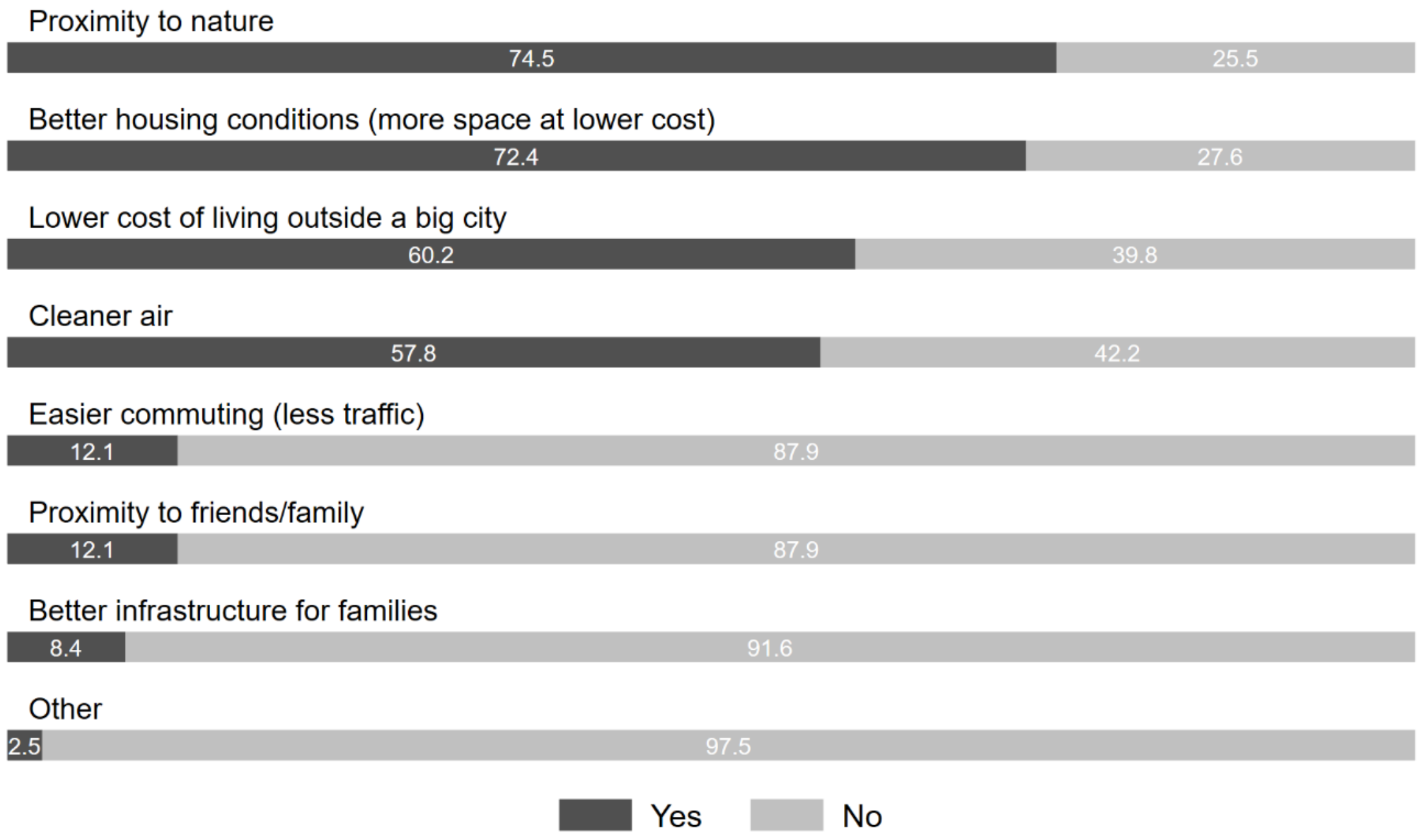

**Figure 5. Key motivations for relocating to the suburbs enabled by remote-work opportunities**

Note: Responses to the question "What would be the main reasons/motivations for your move to the suburbs? Please choose up to three".

Source: *own elaboration based on survey data.*

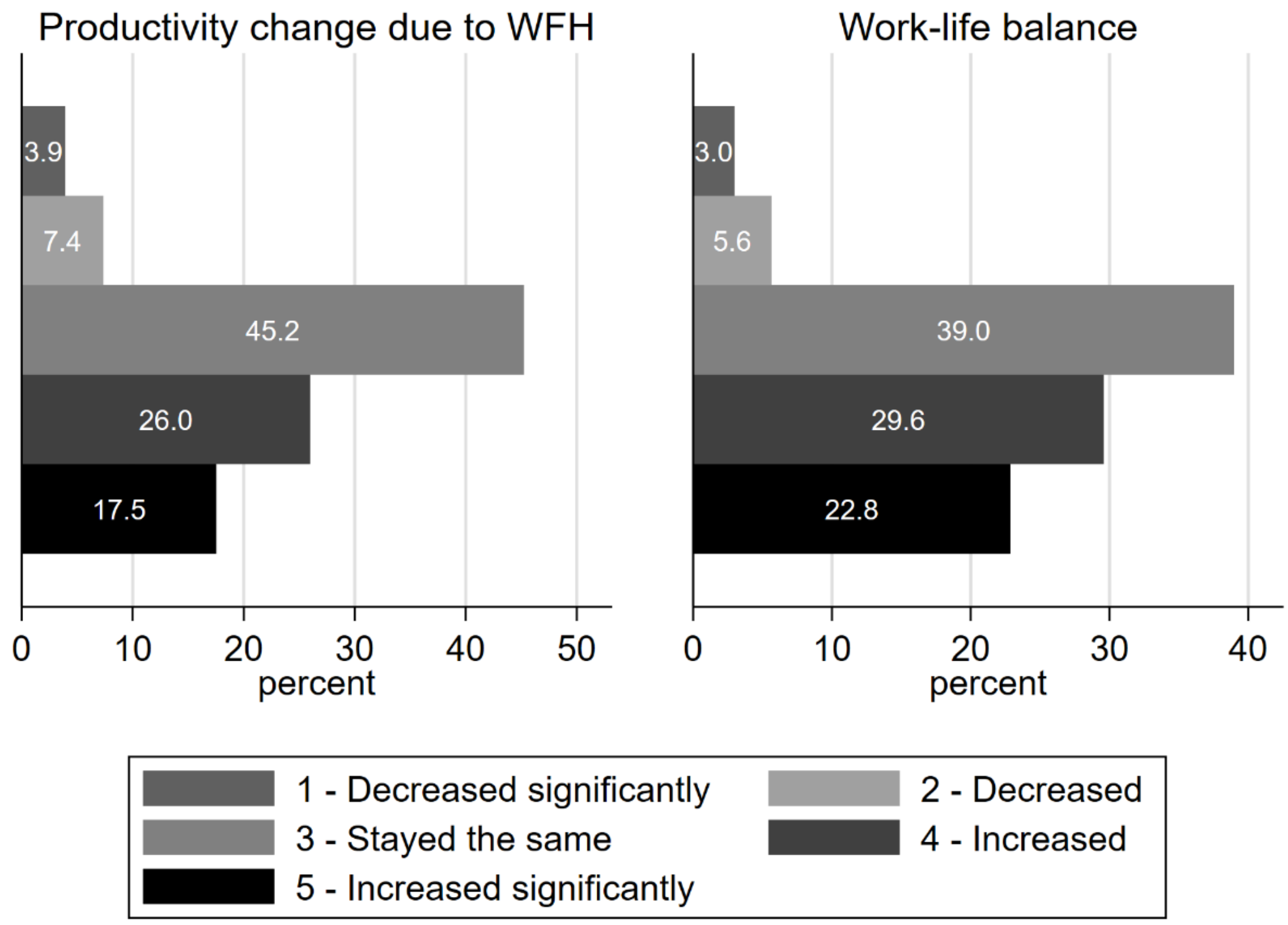

**Figure 6. The main reasons for moving to the suburbs due to WFH opportunities**

Note: Responses to the questions "How has remote work affected your productivity?" and "How has your personal life quality (work-life balance) changed due to remote work?".

Source: *own elaboration based on survey data.*

***Method***

This study examines how socio-economic and job attributes, commuting experiences, and self-reported productivity changes influence the willingness of Polish office workers in large cities to relocate when afforded the option to work from home. Relocation intent was captured with the binary question, "*Would you consider moving to the suburbs if you could WFH?*" (yes/no). Given this dichotomous outcome, we employ a standard logit model to relate the relocation decision to individual and employment characteristics, specified as follows:

$$logit(P) = \log\frac{P}{1-P} = \alpha_1 + \beta X \quad (1)$$

Where $\alpha$ represents the log-odds of the outcome when the predictor $X$ equals 0, and $\beta$ denotes the coefficients estimated for each variable. Logistic regression is a cumulative probability model in which the individual's independent characteristics determine the log odds or logit of belonging to a specific group. The cumulative odds are calculated and then subtracted. This research focuses on the impact of predictor variables on preferences toward relocation rather than on calculating the exact probabilities for each category. As a result, the exact probabilities based on specific predictors of relocation were not analysed. Instead, the emphasis was placed on understanding the relationships between the predictor variables and the tendencies toward relocation preferences, allowing for insights into how changes in the predictors influence these preferences without delving into the precise probability calculations.

Consequently, interpretation centres on the signs and magnitudes of the coefficients and their associated odds ratios, rather than on case-specific probability estimates. A positive coefficient (odds ratio > 1) signifies higher odds of preferring relocation relative to the reference group, whereas a negative coefficient (odds ratio < 1) indicates diminished inclination. For instance, if "*male*" is the baseline for gender, a positive coefficient for "*female*" would imply that women are more likely than men to consider relocating. However, because the model is based on a logistic distribution, the coefficients cannot be interpreted as straightforwardly as in a linear model, where a one-point change in the independent variable directly affects the dependent variable by the coefficient value. Therefore, the odds ratio offers a clearer interpretation: it indicates how much the odds of the outcome change with a one-unit increase in the predictor variable. For continuous variables, this reflects the effect of a one-unit change, while for categorical variables, it shows the shift in odds between different levels of the variable.

A key aspect we sought to underline in this exploratory study is the influence of remote work opportunities on individuals' willingness to relocate. Rather than assessing the practical feasibility of relocation within specific professions, our aim was to capture general preferences and attitudes under the assumption that remote work is a viable option. Given the rapid digital transformation across sectors, we framed the relocation question as a hypothetical scenario to reflect potential, rather than current, work arrangements. This approach allowed us to explore how the prospect of remote work shapes relocation preferences, even among those whose current jobs do not support it.

The analysis explicitly treats the possibility of working from home as a realistic counterfactual, reflecting the accelerating digitalisation of work and its potential to decouple residence from workplace. This framing allows investigation of relocation attitudes even in occupations that may not yet permit remote work. Productivity change[3], rather than self-reported work–life balance, was retained as a predictor to avoid multicollinearity (the two variables correlate at 0.59). To preserve statistical power with a limited sample, educational qualifications were collapsed into broader tiers, and the presence of children under 18 was coded as a single indicator. Collectively, these modelling choices provide a robust platform for evaluating how remote-work opportunities reshape urban workers' residential preferences.

***Results***

Table 2 presents the logistic regression estimates that link respondents' characteristics to their willingness to relocate if remote work becomes feasible. Age emerges as the most consistent predictor: compared with employees younger than 35, the odds of expressing a relocation interest fall steadily across successive cohorts. Specifically, workers aged 35–49, 55–60, 61–65, and 66 or older are, respectively, 34 percent, 49 percent, 72 percent, and 78 percent less likely to consider moving (odds ratios of 0.659, 0.509, 0.283, and 0.220; all $p < 0.05$). In contrast, neither gender nor the presence of children under 18 exerts a statistically discernible influence.

Commuting conditions also matter, but only for certain modes. Individuals who ride to work as car passengers are almost twice as likely as others to contemplate relocation (odds ratio = 1.943, $p < 0.05$). The opposite pattern appears for those grouped under the residual "other" category, generally respondents who already work from home or live within walking distance of the workplace, whose odds of moving drop by roughly 60 percent (odds ratio = 0.410, $p < 0.10$). No systematic association is detected for other transport modes because each mode is represented by a separate dummy rather than by a single reference-based variable.

[3] Questions related to relocation motives and acceptable distances were excluded from further econometric analysis because the sample size was limited to 322 respondents expressing their preferences for relocation.

Organisational position (from administrative staff through owner) and tenure in the current role show no coherent pattern with respect to relocation sentiment. Sector, however, does: private-sector employees are almost twice as likely as their public-sector counterparts to favour relocation when work-from-home is available (odds ratio = 1.920, $p < 0.01$).

Perceptions of productivity exert a further, pronounced effect. Respondents who report even a moderate decline in productivity display lower relocation odds than those who perceive a strong productivity gain, and the deterrent is most acute among those who experience a ‘great decrease’: their odds of favouring relocation plunge to one-fortieth of the reference group’s level (odds ratio = 0.025, $p < 0.01$). By contrast, educational attainment and current residential location fail to reach statistical significance in this model. City size appears to tilt relocation intentions slightly in favour of residents of larger urban areas, yet the coefficients are statistically indistinguishable from zero. The evidence is therefore too weak to assert that metropolitan scale meaningfully shapes the willingness to move.

Taken together, the findings suggest that a preference for relocation in response to remote-work opportunities is chiefly driven by youth, car-passenger commuting, private-sector employment, and perceived productivity improvements, whereas older age, minimal commuting costs, and productivity losses substantially dampen the appeal of moving.

**Table 2. Logistic regression model estimates for relocation preferences**

| Predictors | Coefficient | Odds Ratio | P > z |
|---|---|---|---|
| **Gender** | | | |
| Male | ref | ref | ref |
| Female | -0.012 | 0.988 | 0.953 |
| **Age (from 24 years)** | | | |
| 25-34 years | ref | ref | ref |
| 35-49 years | -0.417* | 0.659* | 0.087 |
| 50-54 years | -0.208 | 0.812 | 0.572 |
| 55-60 years | -0.675* | 0.509* | 0.081 |
| 61-65 years | -1.262** | 0.283** | 0.017 |
| 66+ years | -1.514*** | 0.220*** | 0.010 |
| **Home location** | | | |
| City from 100.000 to 199.000 inhabitants | ref | ref | ref |
| City from 200.000 to 500.000 inhabitants | 0.396 | 1.487 | 0.137 |
| City with more than 500.000 inhabitants | 0.387 | 1.473 | 0.109 |
| **Education** | | | |
| Lower than Master's Equivalent | ref | ref | ref |
| Master's Equivalent | -0.019 | 0.981 | 0.922 |
| **Mode of commuting (dummy variables)** | | | |
| Car Driver | -0.129 | 0.879 | 0.571 |
| Car Passenger | 0.664** | 1.943** | 0.044 |
| Bus/tram/metro | -0.242 | 0.785 | 0.266 |
| Bike | 0.411 | 1.508 | 0.131 |
| Rail | 0.200 | 1.221 | 0.634 |
| Walk | -0.206 | 0.814 | 0.392 |
| Other | -0.891** | 0.410** | 0.041 |
| **Perceived Change in Work Productivity** | | | |
| Greatly Increased | ref | ref | ref |
| Somewhat Increased | 0.260 | 1.297 | 0.358 |
| No Change | -0.693*** | 0.500*** | 0.008 |
| Somewhat Decreased | -0.863** | 0.422** | 0.029 |
| Greatly Decreased | -3.683*** | 0.025*** | 0.001 |

| Predictors | Coefficient | Odds Ratio | P > z |
|---|---|---|---|
| **Children Under 18** | | | |
| Zero | ref | ref | ref |
| One or more | 0.198 | 1.219 | 0.324 |
| **Role in Organisation** | | | |
| Office administrative employee | ref | ref | ref |
| Junior specialist | -0.210 | 0.810 | 0.961 |
| Senior specialist | 0.304 | 1.356 | 0.500 |
| Director/Manger | 0.061 | 1.063 | 0.743 |
| Company owner | 0.288 | 1.334 | 0.364 |
| **Number of years in role** | | | |
| Less than a year | ref | ref | ref |
| 1-3 years | -0.219 | 0.803 | 0.469 |
| 4-7 years | 0.341 | 1.407 | 0.152 |
| 8-10 years | 0.097 | 1.102 | 0.749 |
| More than 10 years | 0.394 | 1.482 | 0.334 |
| **Ownership structure** | | | |
| Public | ref | ref | ref |
| Private | 0.652*** | 1.920*** | 0.004 |
| **Sector** | | | |
| Industry | Ref | Ref | Ref |
| Technology and computer science | -0.193 | 0.825 | 0.655 |
| Finance and banking | -0.296 | 0.744 | 0.518 |
| Education | -0.111 | 0.895 | 0.814 |
| Healthcare | 0.002 | 1.002 | 0.996 |
| Trade and services | -0.069 | 0.934 | 0.866 |
| Transport and logistics | 0.030 | 1.031 | 0.952 |
| Media and entertainment | -0.487 | 0.614 | 0.427 |

| | | |
|---|---|---|
| Number of obs. = 637<br>LR chi2(…) = 124,68<br>Prob. > chi2 = 0.000<br>Pseudo R2 = 0.1423 | Log likelihood = | -379.156 |
| | Hit ratio = | 68.6% |

Note: Asterisks indicate statistical significance levels. *** p<0.01, ** p<0.05, * p<0.1.
Source: own elaboration.

## 4. Discussion

Our results reinforce a broader narrative already emerging in the international literature: when employment becomes location-flexible, the long-standing "centripetal" pull of large cities weakens and the latent appeal of the urban fringe resurfaces (Micek et al. 2025). In the Polish context, where the first post-socialist wave of suburban growth was driven chiefly by housing shortages and a new taste for single-family living (Kajdanek 2011; Palak 2016), telework appears to be supplying a second impulse. Younger cohorts, who elsewhere have proved the most geographically mobile (Kim 2011; Akan et al. 2024), again show the greatest readiness to trade central flats for peripheral space; older owners, by contrast, replicate the age-related residential inertia documented in US and European studies. This age gradient suggests that telework will not homogenise Poland's metropolitan geography but could sharpen generational differences in residential choice.

Perceived changes in productivity provide a complementary lens. Our respondents' pattern mirrors the "productivity-enhanced telecommuters" described by Stefaniec et al. (2022) and aligns closely with US evidence showing that self-assessed efficiency gains are the strongest predictor of suburban moves (Akan et al. 2024). Their enthusiasm accords with the job-demands–resources model, which links well-being to a balance between heightened work–family conflict and resources such as social support, dedicated workspace, and financial security (Galanti et al. 2021; Meyer et al. 2021). Where those resources are present, the flexibility dividend noted by De Haas et al. (2020) outweighs the costs; where they are absent, conflict dominates, echoing the pandemic-era findings of Shamshiripour et al. (2020) and the loneliness mitigated by family and peer support documented by Rubin et al. (2020). Still, self-reports must be treated cautiously: they are prone to optimism bias (Dutcher 2012) and often mask the firm-specific contingencies that Gibbs et al. (2023) document.

The commuting experience adds another layer. Literature on travel satisfaction shows that the burdens of a trip depend on both duration and mode (Jacob et al. 2021; Tao et al. 2023). Our evidence points toward mode as the more salient filter: individuals who lack control during the journey, as car passengers, for instance, appear keenest to reduce its frequency, even at the cost of living farther away. Drivers who can maintain door-to-door accessibility after a move display a more neutral stance, mirroring Pritchard and Frøyen's (2019) argument that suburbanisation in car-centric regions often perpetuates, rather than transforms, automobile dependence. By contrast, drivers, cyclists and walkers show neutral or contradictory signals, echoing the mixed welfare outcomes in the commuting literature: longer trips can harm well-being (Ingenfeld et al. 2019; Tao et al. 2023) yet are willingly accepted if they buy superior housing or careers (Clark et al. 2020) and, when they involve active modes, may even raise life satisfaction (Jacob et al. 2021).

Sectoral differences also matter. The greater mobility of private-sector employees is consistent with research showing faster diffusion of telework practices in market-oriented firms (Durbarry 2021) and underlines the importance of regulatory context: Poland's statutory framework for remote work only stabilised in 2023, leaving public organisations relatively constrained during the study period. That asymmetry is already narrowing. The April 2023 Labour-Code amendment gives public bodies the same legal basis for telework as private firms, and forthcoming EU guidelines on 'digital-ready administration' are likely to accelerate uptake.

Contrary to expectations derived from the suburban surge around Warsaw and Wrocław (Tokarczyk-Dorociak et al. 2018), core-city size itself is not salient, suggesting that factors such as housing costs, amenity packages, or infrastructure deficits – already highlighted by Day and Cervero (2010) and Pobłocki (2021) – now outweigh population scale. Household structure is similarly neutral: the presence of children, long a driver of fringe migration, no longer differentiates intent, an outcome that dovetails with Barlindhaug's (2022) observation that some contemporary families value urban proximity over suburban space.

Overall, these patterns hint at a selective, rather than universal, telework-driven suburbanisation. For planners, the central message is less about an overall surge of outward migration than about a differentiated suburbanisation trajectory shaped by age, workplace culture, commuting constraints, and perceived efficiency gains. Future research should therefore integrate life-course perspectives with institutional and mobility frameworks to gauge how far telework will deepen – or merely redistribute – the spatial imbalances already embedded in Poland's post-socialist metropolis.

## 5. Conclusion and Avenues for Continued Inquiry

Viewed holistically, the Polish case illustrates how the diffusion of remote work is reshaping the relationship between labour markets and residential geography: the spatial tether that once bound knowledge-sector employees to the urban core is loosening, but not in a uniform or easily forecastable manner. Remote work operates less as a one-off "shock" than as a structural shift whose repercussions will unfold over multiple housing-market and business cycles. These early signals of suburban interest, therefore, constitute a leading indicator rather than a settled end-state.

To chart that evolution credibly, future research should move beyond single-wave attitudinal surveys and toward longitudinal designs that trace households across time and space. Linking administrative address records with employer-provided telework data would permit causal tests of whether stated intentions translate into actual moves and, crucially, whether moves endure after pandemic-era policies and incentives fade. A second priority is to embed objective productivity and performance metrics in the analysis; doing so will clarify whether residential dispersion creates durable efficiency gains or merely reflects workers' perceptions in the initial novelty phase of hybrid work.

Comparative work is also essential. Poland's housing tenure structure, transport infrastructure, and regulatory path through pandemic emergency statutes are idiosyncratic; parallel studies in contexts with tighter housing markets or stronger public-sector telework mandates would reveal how generalisable the observed patterns are. Finally, the environmental and equity consequences of telework-driven suburbanisation deserve sustained attention. Modelling the carbon implications of longer discretionary trips, the fiscal impacts on urban tax bases, and the distributional effects on workers whose jobs cannot be performed remotely will help planners design policies that harness the flexibility benefits of WFH without entrenching new spatial inequities.

In short, remote work is not merely shifting where people live; it is opening a fresh research frontier at the intersection of housing economics, labour studies, transport planning, and environmental governance. Multi-disciplinary, data-rich approaches will be required to grasp – and steer – the full arc of this transformation.

**Declaration of generative AI and AI-assisted technologies in the writing process**

During the preparation of this work the author(s) used ChatGPT-4.0 in order to improve the grammar and style of the language only. After using this tool/service, the author(s) reviewed and edited the content as needed and take(s) full responsibility for the content of the publication